\documentclass[12pt]{iopart}

\usepackage{iopams}
\usepackage{graphicx}
\usepackage{dcolumn}
\usepackage{bbm}
\usepackage{hyperref}

\usepackage{tikz}

\usepackage{bbold}

\newcommand{\ket}[1]{|#1\rangle}
\newcommand{\bra}[1]{\langle#1|}

\newcommand{\vev}[1]{\langle#1\rangle}

\renewcommand{\S}{\textsc{s}}
\newcommand{\I}{\textsc{i}}

\newcounter{subequation}[equation]

\newcommand{\eqref}[1]{\hyperref[#1]{(\ref*{#1})}}

\newcommand{\xrightarrow}[1]{\stackrel{#1}{\rightarrow}}

\begin{document}
\title{Effective dimensional reduction of complex systems based on tensor networks}

\author{Wout Merbis$^{1,2}$, Madelon Geurts$^{1,2}$, Cl\'elia de Mulatier$^{1,2,3}$ and Philippe Corboz$^1$}

\address{$^1$ Institute for Theoretical Physics, University of Amsterdam}
\address{$^2$ Dutch Institute for Emergent Phenomena, University of Amsterdam}
\address{$^3$ Informatics Institute, University of Amsterdam}
\ead{w.merbis@uva.nl}

\begin{abstract}
The exact treatment of Markovian models of complex systems requires knowledge of probability distributions exponentially large in the number of components~$n$. 
Mean-field approximations provide an effective reduction in complexity of the models, requiring only a number of phase space variables polynomial in system size. However, this comes at the cost of losing accuracy close to critical points in the systems dynamics and an inability to capture correlations in the system.
In this work, we introduce a tunable approximation scheme for Markovian spreading models on networks based on Matrix Product States (MPS). 
By controlling the bond dimensions of the MPS, we can investigate the effective dimensionality needed to accurately represent the exact $2^n$ dimensional steady-state distribution. We introduce the entanglement entropy as a measure of the compressibility of the system and find that it peaks just after the phase transition on the disordered side, in line with the intuition that more complex states are at the `edge of chaos'.  
We compare the accuracy of the MPS with exact methods on different types of small random networks and with Markov Chain Monte Carlo methods for a simplified version of the railway network of the Netherlands with $55$ nodes. The MPS provides a systematic way to tune the accuracy of the approximation by reducing the dimensionality of the systems state vector, leading to an improvement over second-order mean-field approximations for sufficiently large bond dimensions. 
\end{abstract}


%


\section{Introduction}
Complex systems contain a great number of components, whose nature changes dynamically due to interactions with each other and the environment. Such systems are notoriously hard to study or model, as relevant large-scale properties may \textit{emerge} in seemingly unpredictable ways from the microscopic constituents. In essence, emergence stems from an insensitivity of macroscopic properties of the system to the microscopic behavior of its components. One does not need to know about the microscopic laws of physics to explain and predict the behavior of a system on a large scale. The large scale properties of the system are well described by an \textit{effective model} which only accounts for ensemble averaged (or coarse-grained) observables. This entails an effective dimensional reduction as larger length scales are probed.

A prototypical model of a complex system concerns information spreading on a complex network, where nodes may change their states probabilistically, depending on the state of their neighbor. Such models are relevant for epidemiology, modeling infectious disease or social contagion spreading, and have been widely used in the literature~\cite{PastorSatorras2015,van2008virus,simon2011exact,castellano2009statistical}. On the large scale, such systems are often described using compartmental models~\cite{Anderson1992}, which are mean-field models whose deterministic evolution is described by a coupled set of non-linear ordinary differential equations (ODEs). More refined node-based mean-field approximations on several levels can provide a mesoscopic description~\cite{kiss2017mathematics,van2011n}, taking into account the individual nodes states, and in the case of second order mean-field approximation also pair correlations~\cite{cator2012second}. 

Each level of description has a phase space whose dimensionality depends on the number of states of each node $d$ and the number of components $n$: the stochastic network models require a phase space exponential in system size, of order $\mathcal{O}(d^n)$. Node-based mean-field approximations are of order $\mathcal{O}(d n)$ (first order) or $\mathcal{O}(d^2 n^2)$ (second order), while the deterministic compartmental models can be described in terms of $\mathcal{O}(d)$ differential equations. The node-based mean-field approximations and the compartmental models follow from the collective dynamics of the microscopic Markovian model in a large $n$ limit (see for instance~\cite{merbis2021exact,merbis2023emergent}) and can hence be understood as emergent models on larger scales.

An important question is then: when are the microscopic models, with their potentially exponentially large phase space, accurately described in terms of a lower-dimensional effective model? And related to this question: how can we measure the effective size and hence the compressibility of the microscopic model? Are the mean-field approximations sufficiently complex to capture all the relevant degrees of freedom of the microscopic system? These questions are especially pressing when the system is \textit{away} from the thermodynamic limit of a large number of nodes, where finite size effects play an important role. 

The questions posed above bear a similarity with problems in the area of quantum many-body systems. In this field, many-body Hamiltonians describe the evolution of wavefunctions in an exponentially large Hilbert space. However, whenever the Hamiltonians describe \textit{local} interaction (when nodes/lattice sites interact only with $k \ll n$ nearest neighbors), the low-energy states of the system reside in a much smaller region of Hilbert space, which for one-dimensional systems grows at most polynomially in system size~\cite{poulin2011quantum}. This fact is exemplified by ground states of many-body Hamiltonians which satisfy the \textit{area law} of entanglement: the entanglement entropy between a sub-region $A$ and its complement scales only with the area of this region $\partial A$, and not with its volume. This implies that the main contribution to the quantum correlations in the system come from entanglement across the boundary of the sub-region of interest. Many-body wavefunctions satisfying the area law can be faithfully represented using a lower dimensional representation in terms of tensor networks~\cite{verstraete2006matrix,verstraete2008matrix,orus2014practical}.

\begin{figure}
    \centering
    \includegraphics[width=\linewidth]{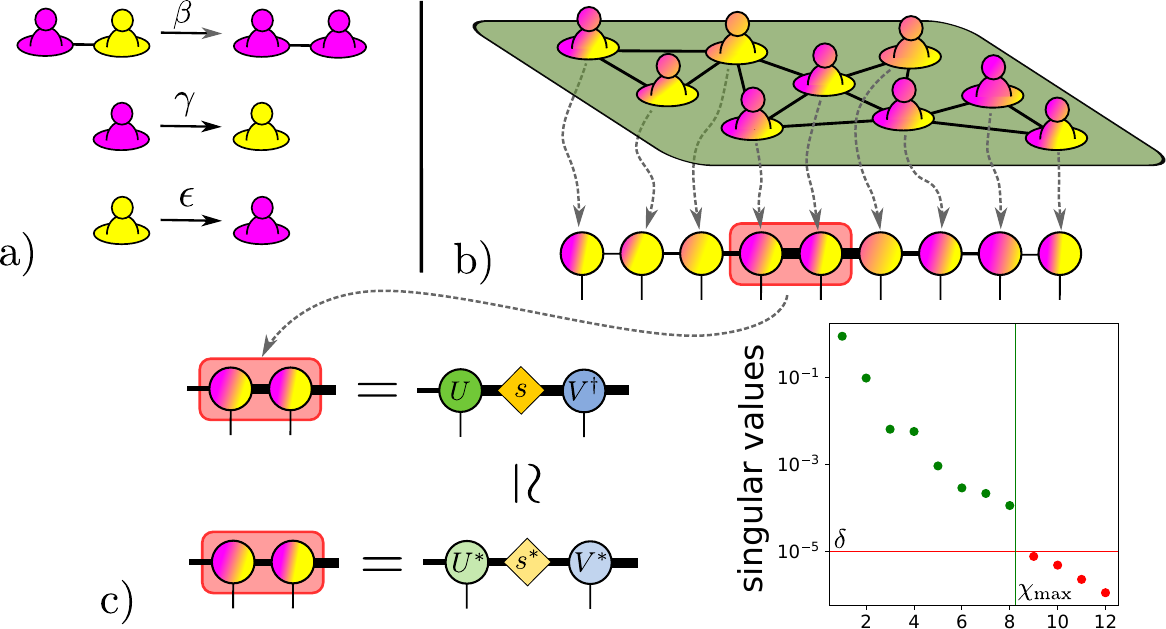}
    \caption{Overview of the MPS approximation method. a) The allowed transitions of the stochastic $\epsilon$-SIS process: connected nodes can transmit an infection (purple) to their susceptible neighbors (yellow) with rate $\beta$, while they recover with rate $\gamma$. Susceptible nodes are spontaneously infected with rate $\epsilon$. b) The probability vector of a network state (top) is mapped to a Matrix Product State (MPS, bottom). The MPS consists of an array of tensors contracted over bond indices. The unconnected (physical) indices represent the two states the node may be in. c) Each bond in the MPS is optimized by performing a singular value decomposition and then truncating the singular values below a threshold $\delta$, or keeping a maximal of $\chi_{\rm max}$ singular values. The plot on the right shows the singular values of this particular bond, illustrating that we keep the largest singular values (in green) in the truncated diagonal matrix $s^*$. In this way a compressed, lower-dimensional, approximation of the state vector can be obtained. }
    \label{fig:overview}
\end{figure}

Tensor networks are networks where nodes represent tensors, and the degree of a node corresponds to the rank of the tensor. Connected edges in a tensor network represent the `contraction' of the connected tensors by matrix multiplication along the corresponding indices. A tensor network may contain open (or dangling) edges, connected only to a single node. These represent indices (or directions) of the tensor which are not contracted with any other tensor, such that the resulting node is vector-valued. Any vector of size $d^n$ (such as the probability vector of a Markovian network model) can be brought into a \textit{Matrix Product State} (MPS) form~\cite{schollwock2011density}, where the vector is decomposed into $n$ rank-3 tensors of size $d*\chi^2$, where $\chi$ denotes the \textit{bond dimension} of the MPS (see the graphical representation in Figure~\ref{fig:overview}b). The $n$ rank-3 tensors are contracted as a linear chain over the bond indices, leaving a $d$ dimensional index for each tensor uncontracted. This results in an object with size of order $\mathcal{O}(n d \chi^2)$, which can give a lower dimensional representation of the original vector if $\chi$ is chosen to be smaller than $d^{n/2}$. Crucially, since the correlations between nodes are now represented by the bonds between the MPS tensors, the tensor network architecture does not need to reflect the underlying complex network on which the dynamics takes place.

The efficiency of the MPS representation hinges on the bond dimension needed to accurately represent the large dimensional vector. Fortunately, one can tune the accuracy of the representation by adjusting the maximal bond dimension. For each bond in the MPS, one can perform a \textit{singular value decomposition} (SVD), where the number and magnitude of singular values correspond to the number and weight of the linear combinations of basis states needed to describe the correlations encapsulated by the bond. 
When the bond dimension scales as $\chi \sim d^{n/2}$, the MPS representation is exact and all basis states are relevant. Usually, however, the number of significant singular values is much smaller, indicating that only a subset of basis states are relevant to describe the larger vector. Hence, the number (and distribution) of significant singular values tells us about the compressibility of the large dimensional state and hence it gives insight into the effective number of degrees of freedom needed to describe the system. Moreover, by truncating the lowest singular values below a fixed threshold $\delta$, or by putting a cap on the maximal number of singular values for each bond $\chi_{\rm max}$, an approximation of the high-dimensional system can be found~\cite{orus2014practical} (see Figure~\ref{fig:overview}c).

While primarily developed in the field of many-body quantum systems, tensor network methods and algorithms, such as the Density Matrix Renormalization Group (DMRG)~\cite{white1992density}, have seen successful applications for stochastic systems on regular lattices (see for instance~\cite{hieida1998application,carlon1999density,Johnson2010,Johnson2014,banuls2019using,causer2022finite,merbis2023efficient}). Recent works have generalized tensor network methods based on Alternating Linear Scheme (ALS) to epidemic spreading models on networks in~\cite{dolgov2024tensor,dolgov2024tensor2}. Tensor network methods applied to non-equilibrium Glauber dynamics on networks have been studied in~\cite{barthel2018matrix,barthel2020matrix} and another recent work is based on matrix product believe propagation algorithms~\cite{crotti2023matrix}. Tensor networks may be efficiently used to compress multivariate functions~\cite{tindall2024compressing} or to efficiently define and sample discrete diffusion models~\cite{causer2024discrete} or two-dimensional spin models~\cite{frias2023collective}.  Other dimensional reductions based on field theory methods to study epidemic spreading using the Doi-Peliti formalism have recently been explored in~\cite{visco2024effective,rojas2024quantum}, see~\cite{del2024field} for a review. An alternative approximation scheme based on cavity master equations on networks was explored recently in~\cite{ortega2022dynamics}.

In this work, we construct and study a Matrix Product State representation of the steady-state distribution of the susceptible-infected-susceptible (SIS) model of epidemic spreading on a complex network. In this model, infected nodes in a network may infect their neighbors with rate $\beta$, while infected individuals recover with a rate $\gamma$. Additionally, we allow for spontaneous infection of susceptible individuals with a rate $\epsilon$, which prevents the system from relaxing to the absorbing healthy configuration (see Figure~\ref{fig:overview}a). This model, sometimes deemed the $\epsilon$-SIS model~\cite{van2012epidemics}, has no known exact solution for the steady-state distribution on generic networks and its complexity depends on the proximity to a critical point $\lambda_c$ in the control parameter $\lambda = \beta/\gamma$. Below this \textit{epidemic threshold}, the infection dies out before it can spread across the network, characterizing an \textit{inactive phase} in the dynamics. Above the threshold, the steady-state distribution contains a non-zero number of infected nodes on average and the infection is said to be \textit{endemic}. For finite networks, the transition between the inactive and the endemic phases cannot be sharply defined due to finite size effects, hence we will look at two points close to criticality. One corresponds to values of $\lambda$ where the variance in infection density is largest, which defines the epidemic threshold in the thermodynamic limit of large $n$. Another interesting point corresponds to values of $\lambda$ where the entropy of the singular value spectrum of the MPS (to be defined below) peaks. We find that this latter point, where the MPS is least compressible, is located more towards the disordered (endemic) side of the phase transition. 

By building the MPS representation of the steady-state, we are able to analyze its compressibility by studying how the number and distribution of relevant singular values changes for various values of $\lambda$ and for various network topologies. This allows us to study the effective dimensionality needed to accurately represent the non-equilibrium steady-state of the Markovian network model, and hence gives insight into when the emergent meso- and macroscopic models are expected to give accurate representations of the systems dynamics. We do this for several small randomly generated graphs, which allows for comparing the accuracy of the MPS with mean-field methods, using the exact Markovian steady-state distribution as ground truth. To push the MPS methods outside of what is possible with exact methods, we construct a MPS representation for a $55$ node network based on the Dutch railway network. In this case we test the accuracy of the MPS against mean-field methods by using Markov Chain Monte Carlo simulations of the stochastic process as a reference.  

This work is organized as follows: In section~\ref{sec:networkSIS} we present the model and the parameters used throughout this work, as well as review the mean-field approximations. Section~\ref{sec:TNmethods} discusses the MPS representation of the probability vector and we define the entanglement entropy here. We furthermore discuss the Matrix Product Operator (MPO) construction for the infinitesimal generator of the network $\epsilon$-SIS Markov process and briefly review the DMRG methods used to obtain the non-equilibrium steady-state distribution. In section~\ref{sec:results} we discuss our results, first for small random networks and then for the graph based on the railway network of the Netherlands. Finally, section~\ref{sec:conclusions} presents our conclusions and comments on possible future work.

\section{The network $\epsilon$-SIS model}
\label{sec:networkSIS}

We will concentrate on a paradigmatic model of contagion spreading on a network, used frequently in the complex systems community: the susceptible-infected-susceptible model with spontaneous infection for each node on a graph, also known as the network $\epsilon$-SIS model~\cite{van2012epidemics}. This is a continuous-time Markov process where each node in a network is in one of two states: infected $\I$ or susceptible $\S$. There are three possible transitions (see Figure~\ref{fig:overview}a):
\begin{itemize}
    \item \textbf{Infection:} $\I + \S \xrightarrow{\beta} \I + \I$  \\
    A node is infected by its infected neighbors with rate $k_\I \beta$, where $k_\I$ is the number of infected neighbors.
    \item \textbf{Recovery:}  $\I \xrightarrow{\gamma} \S$  \\
    Any infected node recovers back to the susceptible state with rate $\gamma$.
    \item \textbf{Spontaneous infection:} $\S \xrightarrow{\epsilon} \I$ \\
    Any susceptible node is spontaneously infected with rate $\epsilon$.
\end{itemize}
There are hence three transition rates $\beta, \gamma$ and $\epsilon$ which parameterize the dynamics, besides the network connectivity. As we are free to scale the time parameter $t$, we may use this freedom to set the recovery rate $\gamma = 1$. The spontaneous infection process can model, for instance, indirect infections, such as by touching infected surfaces or breathing contaminated air. It could also model infections by external sources, for example, by someone who is not accounted for in the studied network. We wish to separate the spontaneous infections from the network transmission dynamics, and hence we set $\epsilon$ to be two orders of magnitude smaller than the recovery rate: $\epsilon/\gamma = 10^{-2}$, for the remainder of this work. The only remaining parameter in the model is then the effective (dimensionless) transmission rate $\lambda = \beta/\gamma$.

The $\epsilon$-SIS process is one of the most basic compartmental spreading models on a network containing a non-trivial steady-state. In comparison, without spontaneous infections, the SIS model contains an absorbing state in which all the nodes are healthy. In that case any finite network will relax to the absorbing state eventually, regardless of the value of $\lambda$. When also $\gamma = 0$, the model reduces to the SI-spreading model on a network, in which case the dynamics is integrable and can be solved exactly in terms of a diagrammatic subgraph expansion \cite{merbis2022logistic}. 
However, the $\epsilon$-SIS model has no known exact solutions for the steady-state distribution on generic networks, although exact results are known for specific, highly symmetric graphs~\cite{simon2011exact,cator2013susceptible,achterberg2022analysis,achterberg2023analytic}.  For large network sizes $n \to \infty$, there is a dynamical phase transition at the epidemic threshold $\lambda_c$, which separates a healthy, inactive phase and an endemic, active phase at late times.

\subsection{Markovian network $\epsilon$-SIS model}

The state of the system is determined in terms of a probability vector $\mathbf{p}(t)$, whose components represent the probability of the network to be in any of its $2^n$ configurations. Without loss of generality, we may write this vector in a basis composed of tensor products of individual state vectors $\ket{i}$:
\begin{equation}\label{probvec}
    \mathbf{p}(t) = \sum_{i_1, \ldots, i_n} p_{i_1 \ldots i_n}(t) \ket{i_1} \otimes \ldots \otimes \ket{i_n}\,,
\end{equation}
where the nodes state basis vectors are defined as:
\begin{equation}
    \ket{i=\S} = \begin{pmatrix}
    1 \\ 0
\end{pmatrix}  \quad \mathrm{and:} \quad \ket{i=\I} = \begin{pmatrix}
    0 \\ 1
\end{pmatrix}  \,.   
\end{equation}
The coefficients $p_{i_1 \ldots i_n}(t)$ in~\eqref{probvec} represent the probability of the network being in configuration $i_1 \ldots i_n$ at time $t$. The time evolution of the probability vector~\eqref{probvec} is given by the master equation
\begin{equation}\label{mstreqn}
\frac{d}{dt} \mathbf{p}(t) = Q(A, \lambda,\epsilon) \cdot \mathbf{p}(t)\,,
\end{equation}
where $Q$ is the transition rate matrix, which depends on the control parameter $\lambda$, as well as the networks adjacency matrix $A$ and the spontaneous infection rate $\epsilon$. We shall assume here that $A$ is fixed, undirected and unweighted, meaning that the transmission rate is equal for each edge and the infection may spread both ways along the edge.\footnote{It is not difficult to relax these assumptions to account for weighted and directed graphs. Time-dependent networks becomes more difficult, however.}

The transition rate matrix can be written in the same tensor product basis as~\eqref{probvec}~\cite{sahneh2013generalized,merbis2021exact,merbis2023emergent}, where it is decomposed into a sum of local operators which act on the individual nodes of the network:
\begin{equation}\label{Qconstruction}
    Q(A, \lambda, \epsilon) = \lambda \sum_{i, j} A^{ij} \hat{q}_{\S\I}^i \hat{n}^j + \sum_{i} (\hat{q}_{\I\S}^i + \epsilon \, \hat{q}_{\S\I}^i)\,.
\end{equation}
Here
\begin{equation}
	\hat{q}_{\S\I} = \left( 
	\begin{array}{cc} 
		-1 & 0 \\ 
		1 & 0 
	\end{array} 
	\right), \quad  
	\hat{q}_{\I\S} = \left( 
	\begin{array}{cc} 
		0 & 1 \\ 
		0 & -1 
	\end{array} 
	\right).
\end{equation}
are the infinitesimal stochastic matrices which flip a single site from the $\S$ state to the $\I$ state and back, respectively.  The operator $\hat{n} = \left(\begin{array}{cc} 0 & 0 \\ 0 & 1 \end{array}\right) = \ket{\I}\bra{\I}$ is the projection matrix on the infected state.  In~\eqref{Qconstruction} the superscript $i$ signifies that this operator acts only on site $i$ in the network. Practically, this implies that, for instance, $\hat{n}^i$ is a tensor product of $(n-1)$ identity matrices, with $\hat{n}$ inserted at site $i$ (and similarly for $\hat{q}^i_{\S\I}$)~\cite{merbis2023emergent}. 

\subsection{Node-based mean-field approximations}
The Markovian model above has a phase space exponentially large in system size, as the probability of being in each of the $2^n$ microscopic configurations is being tracked. A lower-dimensional representation is given by node-based mean-field approximations. Following~\cite{merbis2023emergent}, we may recover the various node-based mean-field limits from the master equation~\eqref{mstreqn}, leading to:
\begin{equation}\label{nodebasedSIS}
	\partial_t \vev{\hat{n}^i (t)} = \lambda \sum_{k=1}^{n} A^{ki} \vev{\hat{n}^k \hat{v}^i(t)} - \vev{\hat{n}^i (t)} + \epsilon (1- \vev{\hat{n}^i(t)})\,.
\end{equation}
Here $\vev{\hat{n}^i(t)}$ is the expectation value that node $i$ is infected and $\hat{v}$ is the projection operator on the susceptible state $\hat{v} = \ket{\S}\bra{\S}$, such that $\vev{\hat{v}^i(t)}$ represents the expectation value for node $i$ to be susceptible. We have also used the constraint that $\vev{\hat{v}^i} = 1 - \vev{\hat{n}^i}$ at all times.

The expectation value for the nodes $i$ and $k$ to form an $\S\I$ pair is denoted as $ \vev{\hat{n}^k \hat{v}^i(t)} $ and its dynamical evolution depends on the presence of other pair and triplet expectation values. The first-order mean-field representation is obtained by assuming the absence of node correlation, such that $ \vev{\hat{n}^k \hat{v}^i(t)}  = \vev{\hat{n}^k(t)} \vev{\hat{v}^i(t)}$ and~\eqref{nodebasedSIS} becomes:
\begin{equation}\label{firstorderMFT}
	\partial_t \vev{\hat{n}^i (t)} = \lambda \sum_{k=1}^{n} A^{ki} \vev{\hat{n}^k(t)}(1- \vev{\hat{n}^i(t)}) - \vev{\hat{n}^i (t)} + \epsilon (1- \vev{\hat{n}^i(t)})\,,
\end{equation}
The dynamics is now effectively reduced to $n$ coupled bilinear ODEs.
In actuality, the dynamical evolution of the system does induce node correlations and hence will deviate from the first-order mean-field approximations. In order to create a more accurate mean-field representation, the correlations between nodes may be tracked, leading to an additional set of equations for $\vev{\hat{v}^i \hat{v}^j}, \vev{\hat{n}^i \hat{v}^j} $ and $\vev{\hat{n}^i \hat{n}^j}$. Working this out, using the general methodology presented in~\cite{merbis2023emergent}, gives:
 \begin{eqnarray}
\hspace{-2cm}	\partial_t  \vev{\hat{v}^k \hat{v}^i} & = & \, - \lambda \sum_{l=1}^n  \left(  A^{lk} \vev{\hat{n}^l \hat{v}^k \hat{v}^i} + A^{li} \vev{ \hat{v}^k \hat v^i \hat{n}^l} \right) + \vev{\hat{n}^k \hat{v}^i} +  \vev{\hat{v}^k \hat{n}^i} - 2\epsilon \vev{\hat{v}^k\hat{v}^i} \,, \label{vvpairs} \\
\hspace{-2cm}  	\partial_t  \vev{\hat{n}^k \hat{v}^i} & = & \, \lambda \sum_{l=1}^n  \left(  A^{lk} \vev{\hat{n}^l \hat{v}^k \hat{v}^i} - A^{li} \vev{ \hat n^k \hat v^i \hat{n}^l} \right) + \vev{\hat{n}^k \hat{n}^i} -  (1+\epsilon) \vev{\hat{n}^k \hat v^i} + \epsilon \vev{\hat{v}^k \hat{v}^i}  \,, \label{nvpairs}\\
\hspace{-2cm}   \vev{\hat{n}^i\hat{n}^j}  & = &\,  1 - \vev{\hat{v}^i\hat{v}^j} - \vev{\hat{n}^i\hat{v}^j} - \vev{\hat{v}^i\hat{n}^j}. \label{nnconstraint}
\end{eqnarray} 
Since we only need to track correlations for nodes $k$ and $i$ if they are connected, equation~\eqref{vvpairs} represents $m$ independent equations and~\eqref{nvpairs} are $2m$ equations, where $m$ is the number of edges present in the network.
We may now close this set of equations at the second order, by representing triples in terms of pairs and single node expectation values. There are several ways to do so, however, in this work we will follow the pair closure from~\cite{kiss2017mathematics}. This amounts to representing triples of the form $\vev{X^i X^j Y^k} \simeq \frac{\vev{X^i X^j}\vev{X^j Y^k}}{\vev{X^j}} $ whenever the nodes $i,j,k$ do not form a triangle. When $i,j,k$ are part of a connected triangle, the same quantity is replaced by $\frac{\vev{X^i X^j}\vev{X^j Y^k}\vev{X^i Y^k}}{\vev{X^i}\vev{X^j}\vev{Y^k}}$.
After performing the closure at the second level, we have $n+ 3m$ coupled ODEs~\cite{kiss2017mathematics}. 

In principle, the approximation can be improved further by formulating closures at the quartic or higher levels, which comes at the expense of introducing more differential equations. In this work, however, we wish to explore another way to approximate the complex spreading process, based on a different way to reduce the dimensionality of the microscopic Markovian probability vector. We will use techniques based on tensor networks to do so.

\section{Tensor network methodology}
\label{sec:TNmethods}

We will now provide details on the MPS representation of the $\epsilon$-SIS steady-state distribution and the way we construct the Markovian transition rate matrix~\eqref{Qconstruction} as a Matrix Product Operator (MPO). We also briefly describe the density matrix renormalization group (DMRG) algorithm used to find the non-equilibirium steady-state distribution as an MPS, although these are by now standard methods covered in detail in the tensor network literature (see for instance the reviews  ~\cite{verstraete2008matrix,schollwock2011density,orus2014practical}).

\subsection{Matrix Product States}

The state vector $\mathbf{p}$ in~\eqref{probvec} is a $2^n$ dimensional vector, which can be reshaped into a rank $n$ tensor where each of the tensors 2-dimensional axes corresponds to the 2-dimensional physical state space for each node. Graphically, we may depict this as a box with $n$ open legs:
\begin{equation}\label{ptensor}
    p_{i_1 \ldots i_n} =
\begin{array}{c}
\begin{tikzpicture}
  \draw[thick, fill=orange!40, rounded corners=8pt] (0,0) rectangle (4,1);
  \draw[thick] (0.5,-0.5) -- (0.5,0); 
  \node at (0.5,-0.75) {$i_1$};      
  \draw[thick] (1,-0.5) -- (1,0); 
  \node at (1,-0.75) {$i_2$};      
  \node at (1.9,-.35) {$\ldots$};
  \draw[thick] (2.75,-0.5) -- (2.75,0); 
  \node at (2.75,-0.75) {$i_{n-1}$};  
  \draw[thick] (3.5,-0.5) -- (3.5,0); 
  \node at (3.5,-0.75) {$i_n$};      
\end{tikzpicture}
\end{array} \,.
\end{equation}
Any such vector may be expressed as a Matrix Product State, which is a one-dimensional array of tensors contracted over internal bond indices~\cite{schollwock2011density}. This can be done by a sequence of \textit{singular value decompositions} (SVDs), which implement a Schmidt decomposition of each site's basis states with the remainder of the tensor.  Although this procedure is inefficient and hardly used in practice, it is illustrative to explain the basic idea of the MPS compression. First, one reshapes the tensor as a $(2, 2^{n-1})$ matrix and performs the SVD to write $p_{i_1 \ldots i_n} = \sum_k U_{i_1 k} s_k V_{k i_2 \ldots i_n}$, where $U$ is a $(2,\chi)$ dimensional unitary matrix, $\mathbf{s}$ a vector of $\chi$ singular values and $V$ is a $(\chi, 2^{n-1})$-dimensional matrix. The unitary matrices $U$ and $V$ can be thought of as implementing a basis transformation on the tensor $p$. Now, $\chi$ linear combinations of basis vectors are weighted by the singular values across the bond separating the first node with the rest of the system. This allows us to identify which linear combinations of basis states are most relevant to describe the correlations between the first node and the rest of the system, as these will correspond to the largest singular values in $\mathbf{s}$.

Continuing this process of SVDs for each subsequent bond, systematically multiplying the singular values into the right unitary matrix $V$ as we go along, allows us to write the tensor~\eqref{ptensor} as:
\begin{eqnarray} \label{pmps}
    p_{i_1 \ldots i_n} = &  \sum_{\chi_1, \ldots \chi_{n-1}} (U_1)_{i_1 \chi_1} (U_2)^{\chi_1}_{i_2 \chi_2} \ldots (U_{n-1})^{\chi_{n-2}}_{i_{n-1} \chi_{n-1}} (\mathbf{s}V_n)^{\chi_{n-1}}_{i_n} \\
    = & {\small
\begin{array}{c}
\begin{tikzpicture}
\node[shape=circle,draw=black,fill=green!40] (A1) at (0,0) {$U_1$};
\node[shape=circle,draw=black,fill=green!40] (A2) at (1.5,0) {$U_2$};
\node[shape=circle] (dots) at (3,0) {$\ldots$};
\node[shape=circle,draw=black,fill=green!40] (AN) at (5,0) {$\mathbf{s}V_n$};
\node[shape=circle] (g1) at (0,-1) {};
\node[shape=circle] (g2) at (1.5,-1) {};
\node[shape=circle] (gN) at (5,-1) {};
\path [-,line width=2pt] (A1) edge node[above] {$\chi_1$} (A2);
\path [-,line width=2pt] (A2) edge node[above] {$\chi_2$} (dots);
\path [-,line width=2pt] (AN) edge node[above] {$\chi_{n-1}$} (dots);
\path [-] (A1) edge node[left] {$i_1$} (g1);
\path [-] (A2) edge node[left] {$i_2$} (g2);
\path [-] (AN) edge node[left] {$i_n$} (gN);
\end{tikzpicture}
\end{array} }    
\end{eqnarray}
Here, the size of the bonds $\{\chi_1, \ldots \chi_{n-1} \}$ represent the number of linear combinations of basis states needed to represent the correlations between the left and right sides of each bond. If all singular values are kept, the construction~\eqref{pmps} is exact, but the bond dimensions typically will grow exponentially with each bond in the MPS.
Therefore, an approximation of the larger tensor can be made by truncating the singular value spectrum. Practically, this can be done in two ways. Either the singular values below a cutoff threshold $\delta$ are disregarded, or only a maximal number of singular values $\chi_{\rm max}$ are kept. In practice, we will use both these methods simultaneously, where $\chi_{\rm max}$ will be kept relatively large ($\sim 300$) such that it only comes into effect when the state is very complex and the computation becomes more demanding.  

Effectively, truncating the bond dimension implements a dimensional reduction of the large tensor~\eqref{ptensor}, from the original $2^n$ components, to the MPS, which contains $\mathcal{O}(n \chi^2)$ components. A measure of the compressibility (and hence a proxy for the complexity) of the state we will use throughout the text is the entropy of the singular value spectrum across the largest bond~\cite{bakker2024operator}. More precisely, and in analogy to the quantum mechanical case, we define the \textit{entanglement entropy} of a bond to be the Shannon entropy of the squares of the singular values, normalized to form a probability distribution:
\begin{equation}\label{See}
    S_{\rm ee}(\mathbf{s}) = - \sum_{k = 1}^{\chi} \frac{s_k^2}{|\mathbf{s}|^2} \log \frac{s_k^2}{|\mathbf{s}|^2},
\end{equation}
where $|\mathbf{s}|$ denotes the Euclidean norm of the singular values. The entanglement entropy measures how broad the singular value spectrum is. It vanishes when only one singular value is non-zero. In this case the bond may be represented as a direct product of the left and right sides and there are no correlations between the two sides of the bond. If all bonds have dimension 1, the MPS is in a direct product state and all nodes are uncorrelated. This is the case when the system is in any microscopic configuration with probability 1, but also when each node is statistically independent of its neighbor's state. Hence, the mean-field steady-state solution to~\eqref{nodebasedSIS} can be represented as a direct product state. The entanglement entropy is maximal when all singular values are significant and of the same size. In this scenario, there is no compression possible and all basis states are equally relevant to describe the state of system. The aim of this work is to explore the accuracy and efficiency of this way of compressing the steady-state distribution of the Markovian $\epsilon$-SIS model on complex networks.

\subsection{Matrix Product Operator for the $\epsilon$-SIS generator on a network}
We are interested in obtaining an MPS representation for the steady-state vector $\mathbf{p}_\infty$ of the $\epsilon$-SIS model on a complex network. This vector solves $Q \cdot \mathbf{p}_\infty = 0$ and hence it can be obtained as the (normalized) leading right eigenvector of the transition rate matrix $Q$. Below, we will compute this vector in two ways. One is to use exact diagonalization, where one explicitly constructs the $2^n$ dimensional matrix $Q$ and numerically finds the leading eigenvector. This way is obviously inefficient for large networks, as $Q$ grows exponentially.
Another way to obtain the steady-state distribution is to use the density matrix renormalization group (DMRG) algorithm~\cite{white1992density,schollwock2011density,orus2014practical}. This allows one to search the space of MPSs for a representation close to the exact solution. The error in the approximation is then related to the amount of singular values truncated, as we will demonstrate below. The DMRG method is well documented in various works~\cite{schollwock2011density,orus2014practical} (see~\cite{merbis2023efficient} for an adaptation to stochastic systems), and we refer to there for more details on the implementation. Here we will only briefly explain the algorithm and focus more on the construction of the Markov transition rate matrix~\eqref{Qconstruction} as a Matrix Product Operator (MPO).

In the DMRG algorithm, one starts with an Ansatz for the probability vector $\mathbf{p}_\infty$ already in MPS form, let us call this $\mathbf{p}_{\rm mps}$. One can then sequentially improve $\mathbf{p}_{\rm mps}$, by optimizing each tensor to minimize (the $L^2$) expectation value of $Q$:
\begin{equation}
\vev{Q}_2 = \mathbf{p}_{\rm mps}^T Q \mathbf{p}_{\rm mps}    
\end{equation}
In order to do so, one first writes $Q$ in an MPO form, to allow for easy (node-wise) contraction with the MPS. As shown below, this can be done exactly. Then,  $\vev{Q}_2$ is optimized by sequentially updating the tensors for two nodes at the same time, after which the two nodes are split into two separate tensors using an SVD. The double node update has as advantage that we can dynamically adjust the bond dimension between the nodes, by truncating singular values below the threshold $\delta$, or only considering the largest $\chi_{\rm max}$ singular values.

The DMRG algorithm requires an MPO representation of the transition matrix $Q$. In order to construct this, we first create an MPO with bond dimension 2 to implement the recovery and spontaneous infection transitions for each node:
\begin{equation}
    Q_1 = L \cdot M \cdot M \ldots M \cdot R\,,
\end{equation}
with
\begin{eqnarray} \nonumber
    L  & = & \left(\begin{array}{cc}
        \hat{q}_{\I\S} + \epsilon \hat{q}_{\S\I} & \mathbb{1}
    \end{array}\right) \,,  \qquad
    M  = \left(\begin{array}{cc}
        \mathbb{1} & 0 \\
        \hat{q}_{\I\S} + \epsilon \hat{q}_{\S\I} & \mathbb{1}
    \end{array} \right)\,,  \\
    R & = & \left(\begin{array}{cc}
        \mathbb{1} \\
        \hat{q}_{\I\S} + \epsilon \hat{q}_{\S\I}
    \end{array}\right)\,.
\end{eqnarray}
We then add to this an MPO for each edge in the network connecting the nodes $i$ and $j$: 
\begin{equation}
\hspace{-2cm}    Q_{\rm edge}^{ij} =  \mathbb{1} \cdot \ldots \cdot  \mathbb{1} \cdot \left(\begin{array}{cc}
       \hat{n} & \lambda \hat{q}_{\S\I} 
    \end{array}\right)_i  \cdot 
    \left(\begin{array}{cc}
        \mathbb{1} & 0 \\
        0 & \mathbb{1}
    \end{array}\right) \cdot \ldots 
    \cdot 
	\left(\begin{array}{cc}
        \mathbb{1} & 0 \\
        0 & \mathbb{1}
    \end{array} \right)
    \cdot
    \left(\begin{array}{cc}
        \lambda \hat{q}_{\S\I}  \\
        \hat{n}
    \end{array}\right)_j \cdot  \mathbb{1} \cdot \ldots \cdot \mathbb{1} .
\end{equation}
Here the subscripts $i$ and $j$ label the position in the MPO where the nodes $i$ and $j$ are placed. The final MPO $Q = \sum_{ij} A^{ij} Q^{ij}_{\rm edge} + Q_1 $ then has a maximal bond dimension $\chi_{\rm MPO} = 2+2m$, where $m$ is the number of edges in the network. Using a series of singular value decompositions, we may compress the MPO and reduce this bond dimension further. 

\begin{figure}
    \centering
    \includegraphics[width=0.9\linewidth]{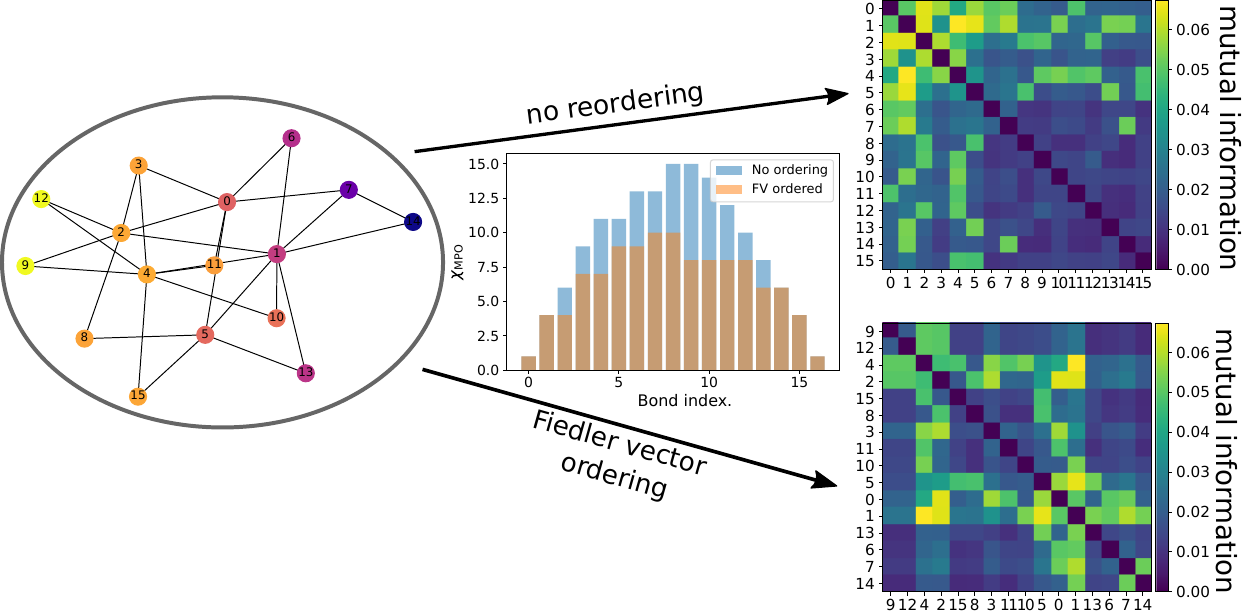}
    \caption{A randomly selected Barab\'asi-Albert network with $16$ nodes, displayed on the left with nodes labeled by the original ordering and colored by the Fiedler vector ordering. We compute the MPS representation of the steady-state once without reordering the nodes and once with an ordering based on the first Fiedler vector. As illustrated in the middle, the MPO bond dimensions are decreased by reordering. On the right we see that the Fielder vector ordering leads to a concentration of the mutual information on the diagonal, meaning that more highly correlated nodes are placed closer to each other in the MPS.}
    \label{fig:Fiedler}
\end{figure}

The above construction is sensitive to the specific ordering of nodes in the MPO/MPS Ansatz, as the nodes of the complex network on which the dynamics takes place are represented by tensors organized in a one-dimensional chain. A more efficient representation will have nodes which are highly correlated placed closer together in the MPS ordering. This problem is reminiscent of the ordering of orbitals in tensor network computations for quantum chemistry~\cite{legeza2003optimizing,barcza2011quantum,szalay2015tensor}. There, the optimal ordering is found by minimizing the distance between orbitals in the MPS configuration based on the mutual information (MI). A $n \times n$-dimensional graph Laplacian is constructed based on the MI, whose Fiedler vector~\cite{fiedler1973algebraic} defines a one-dimensional reordering of the graph which minimizes the expected square distance between nodes~\cite{atkins1998spectral}. Here, since mutual information between nodes is mediated through the contact network, we order the nodes according to the Fiedler vector of the graph Laplacian directly. As we will show below (see Figure~\ref{fig:bonddimmpo}), we found that this drastically reduces the bond dimension as compared with a random ordering, which reduces the computation time. As illustrated in Figure~\ref{fig:Fiedler}, the Fiedler vector ordering ensures highly correlated nodes (as measured by their mutual information) are placed closer together in the MPS representation.

\section{Results}
\label{sec:results}
In this section we provide several results on the accuracy and efficiency of the MPS representations of the $\epsilon$-SIS steady-state distribution. We focus first on small random networks, such that we may compare the MPS to mean-field approximations using the exact solution as ground truth. We then compute the MPS representation of a simplified version of the railway network of the Netherlands with 55 nodes and 83 edges and compare its accuracy to Markov chain Monte Carlo (MCMC) methods.

\subsection{Small random networks}

\begin{figure}
    \centering
    \includegraphics[width=0.75\linewidth]{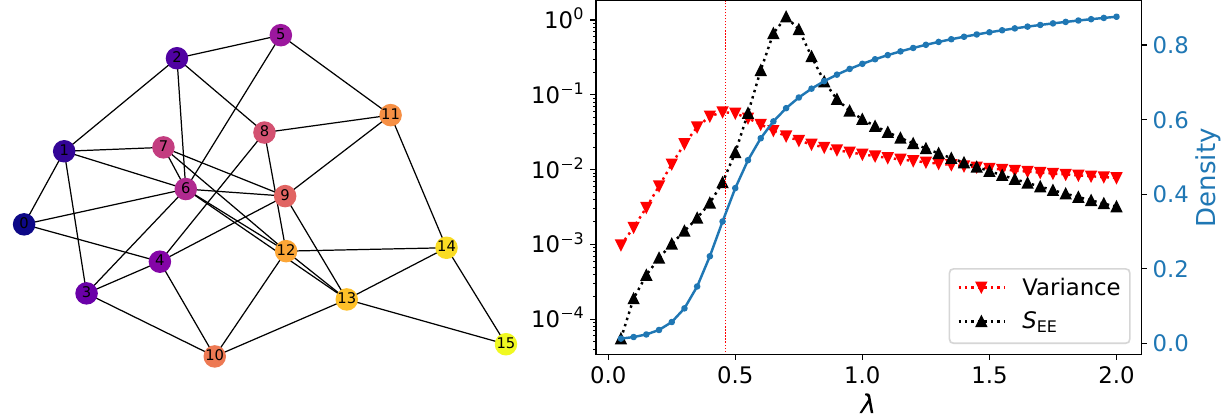}
    \caption{The density (blue, right axis), variance in the density (red, left axis) and the entanglement entropy (black, left axis) for a randomly generated Erd\"os-R\'enyi network with $16$ nodes. The network used is shown on the left and colored and labeled according to its first Fiedler vector. The variance in density peaks at a lower $\lambda$ value compared to the entanglement entropy, indicating that the state is less compressible on the endemic side of the phase transition. These results were obtained with $\delta = 10^{-10}$.}
    \label{fig:var_vs_ee}
\end{figure}

First we focus on small, randomly generated graphs. We choose 4 different types of random graph generators, which are able to produce random graphs which are comparable in terms of the number of edges, but differing in distribution of those edges. This allows us to compare different graph topologies, whose MPOs will have similar bond dimensions and hence are comparable in terms of computational complexity.
The Erd\"os-R\'enyi (ER) networks~\cite{erdos1960evolution} are random networks where all edges are independently present with probability $p=0.25$. The Barab\'asi-Albert (BA) graphs~\cite{barabasi1999emergence} are generated with a preferential attachment scheme resulting in the presence of hubs. Every new node is generated with $m=2$ edges, which are preferentially attached to high degree nodes. The Watts-Strogatz (WS) generator~\cite{watts1998collective} creates networks with a small-world property, where edges from a regular ring lattice with $k=4$ nearest neighbor connections are rewired randomly with probability $p=0.25$. Finally, the Stochastic Block Model (SBM)~\cite{holland1983stochastic} has the nodes divided into two communities of equal size, and the algorithm generates random links within communities with probability $p = 0.4$, while links connecting nodes from different communities are created with probability $q=0.1$. 

To start this section, we first briefly illustrate results for a single random Erd\"os-R\'enyi graph with $n=16$ nodes in Figure~\ref{fig:var_vs_ee}. We plot the density $\rho$, defined as the average number of infected nodes per site, the variance in the density and the entanglement entropy across the middle bond as a function of transmission rate $\lambda$. In the thermodynamic limit, the density becomes the order parameter for the phase transition between the inactive and the endemic phase. When the density starts to increase, the variance peaks, which is taken as a sign that the system is at the critical point (vertical line). It is interesting to note that the entanglement entropy defined as~\eqref{See} peaks \textit{after} the variance peaks. This indicates that the state of the system is more complex \textit{on the disordered side of the phase transition}. Note that the value of the entanglement entropy will depend on the ordering chosen for the MPS, but the fact that it peaks \textit{after} the phase transition point is independent of ordering. We will use this insight to analyze the accuracy of the MPS and the behavior of the entanglement entropy at four different points: in the inactive regime ($\lambda = 0.1$), close to where the variance peaks ($\lambda = 0.55$), close to where the entanglement entropy peaks ($\lambda = 0.85$) and in the endemic regime ($\lambda= 2$).\footnote{Note that these values of $\lambda$ were chosen such that on average, the networks state would be close to the desired regime, but in reality each network will have a slightly different value of $\lambda$ where the variance and entanglement entropy peaks. }

\begin{figure}
    \centering
    \includegraphics[width=\linewidth]{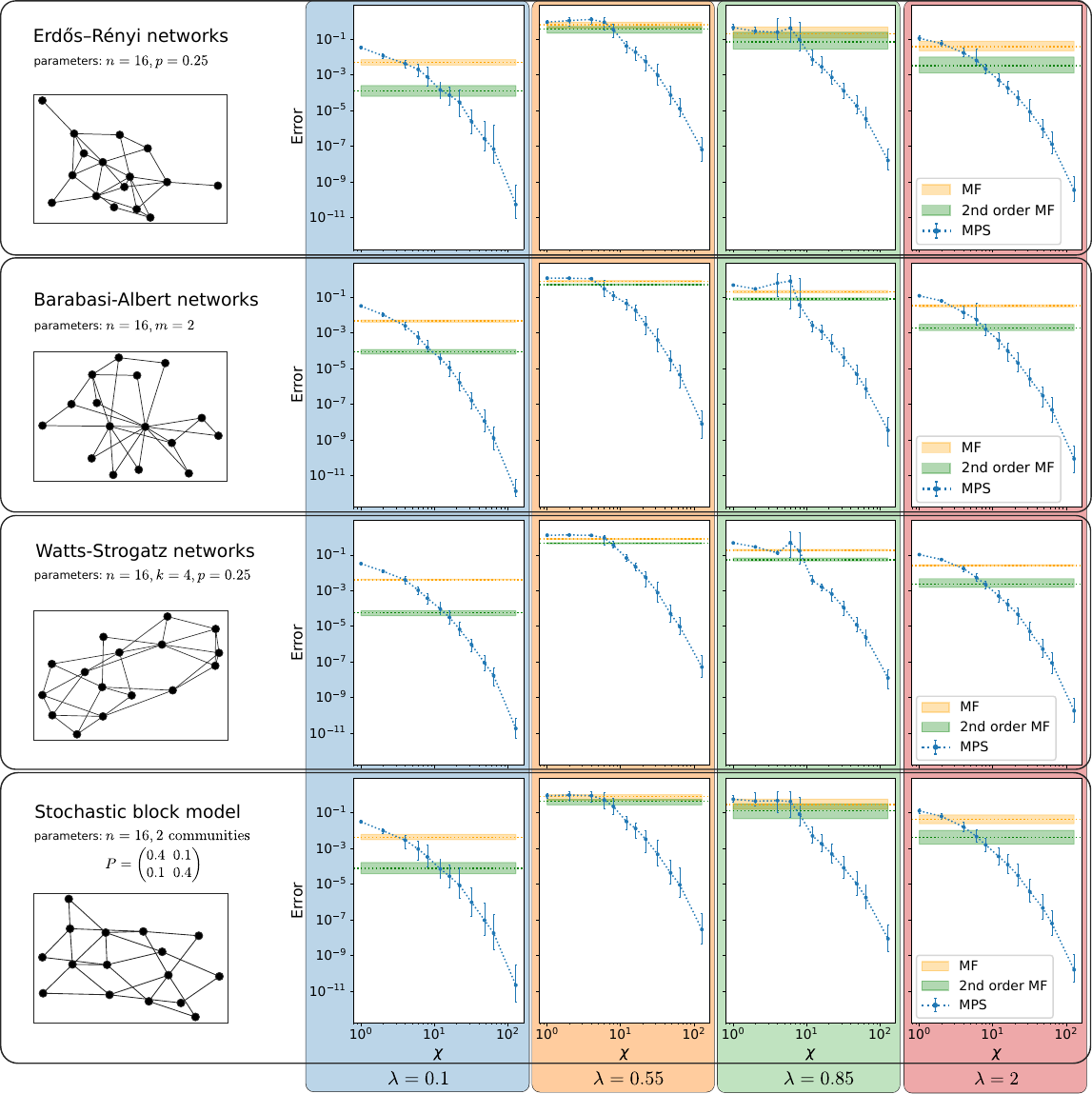}
    \caption{Comparing the accuracy of estimators for local densities $\vev{\hat{n}^i}$ using the MPS representation as a function of maximal bond dimension $\chi$ with the mean-field approximations for four random graph generators. We pick four values of $\lambda$, such that the graphs are in the inactive regime ($\lambda = 0.1$), close to maximal variance ($\lambda = 0.55$), close to maximal entanglement entropy ($\lambda = 0.85$) and in the endemic regime ($\lambda = 2$). The error is defined as the Euclidean distance between the approximation of the single node expectation values and those computed from exact diagonalization.}
    \label{fig:meanfield_comparison}
\end{figure}

\subsubsection{Mean-field comparison}
The MPS method provides a way to tune the accuracy of the approximation, depending on the number of singular values which are kept. By truncating the bonds of the MPS, we are performing a dimensional reduction of the probability vector representing the non-equilibrium steady-state, instead of making a mean-field assumption. Still, we can compare the accuracy of the dimensional reduction with that of the mean-field approximations. This is illustrated in Figure~\ref{fig:meanfield_comparison} for small networks (with $n=16$) generated by the four different graph generators explained above. As ground truth, we construct the $2^n$ dimensional exact transition matrix for each graph and find its leading right eigenvector (using the implicitly Restarted Lanczos Method from the SciPy sparse linear algebra package). Using this exact solution, we compute the $n$-dimensional vector of single node expectation values $\vev{\hat{n}^i}_{\rm exact}$. The error for the various approximations is then estimated by computing the Euclidean distance between $\vev{\hat{n}^i}_{\rm exact}$ and the individual node expectation values computed from the approximation. 

For each network, we compute the mean-field and second order mean-field approximations by setting the time derivatives in~\eqref{firstorderMFT} and~\eqref{vvpairs}-\eqref{nvpairs} to zero and solving for the single node expectation values. These are depicted in Figure~\ref{fig:meanfield_comparison} by the orange and green bars, respectively. The variance in the answer is a result of averaging over 100 different graphs, which all give slightly different expectation values for the same value of $\lambda$. We then compute the same expectation values using the MPS method, while varying the maximal bond dimension from 1 to 128. When choosing $\chi_{\rm max} = 1$, the DMRG algorithm \textit{does not} converge to the mean-field solution, even though the mean-field answer is expressible as a $\chi=1$ MPS. Furthermore, we observe that starting the DMRG optimization on the mean-field solution will cause it to move away from it. In fact, the MPS is generally less accurate than mean-field when $\chi_{\rm max} = 1$, but when increasing the maximal bond dimension, the MPS accuracy quickly improves. Choosing a bond dimension comparable to the system size $\chi \sim \mathcal{O}(n)$ generally leads to an improved accuracy as compared with the second-order mean-field approximation. For even larger bond dimensions, the MPS becomes more accurate than second-order  mean-field theory by several orders of magnitude. 

The accuracy of the MPS for small bond dimensions depends on the region of parameter space, i.e. whether $\lambda$ is tuned to be in the inactive or endemic regime, close to criticality or close to maximal entanglement entropy. In the inactive and endemic regimes, low bond dimensions already give fairly good approximations, although in these regions mean-field theory is generally also much better.  
Near criticality, corresponding to the middle two columns of Figure~\ref{fig:meanfield_comparison}, the MPS performs poor when the bond dimension is too small. For $\lambda = 0.85$ the large variance is due to convergence issues, indicating that the bond dimension chosen is too small to accurately represent the state at this point. However, when $\chi$ is sufficiently large to capture all relevant singular values, the MPS approximation suddenly becomes better than second-order mean-field theory for all graph types. Interestingly, in this region the mean-field accuracy is considerably worse.

\subsubsection{Entanglement entropy as measure of compressibility}

\begin{figure}
    \centering
    \includegraphics[width=\linewidth]{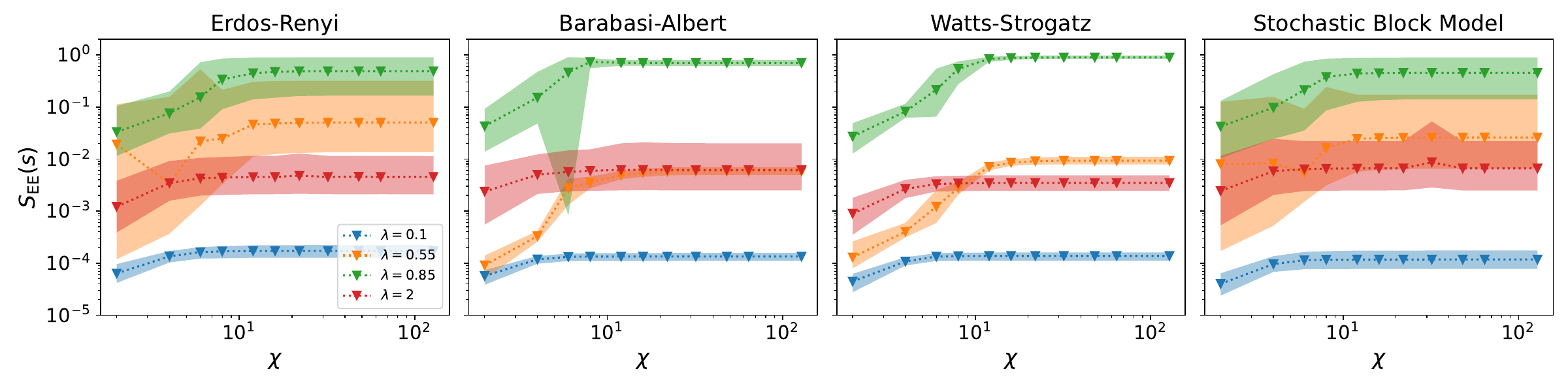}
    \caption{The entanglement entropy as defined in~\eqref{See}, averaged over 100 realizations of random graphs generated by four graph generators (with parameters given in Fig~\ref{fig:meanfield_comparison}). We observe that the entanglement entropy converges for bond dimensions of the order of $n$, which is also where the accuracy starts to improve over the second order mean-field approximation. }
    \label{fig:entanglemententropy}
\end{figure}

The point where the MPS becomes accurate can be found by looking at the entanglement entropy~\eqref{See} across the largest bond in the MPS. In Figure~\ref{fig:entanglemententropy} we plot the $S_{\rm EE}(s)$ as a function of the maximal bond dimension $\chi$, averaged over the 100 randomly generated networks for each of the four types described above. It shows how the entanglement entropy saturates when the bond dimension is increased. The improvement in accuracy compared to the mean-field results, coincides with the values of $\chi$ where entanglement entropy saturates. The large variance in $S_{\rm EE}$ for low bond dimensions when $\lambda = 0.55$ or $0.85$ indicates that an insufficient number of singular values are kept to provide a good approximation. In the endemic and inactive regimes, the entanglement entropy grows monotonically, indicating that in these cases, an insufficient number of kept singular values may still provide decent approximations. This is corroborated by Figure~\ref{fig:meanfield_comparison}, where the error drops smoothly as a function of bond dimension.  

\begin{figure}
    \centering
    \includegraphics[width = \textwidth]{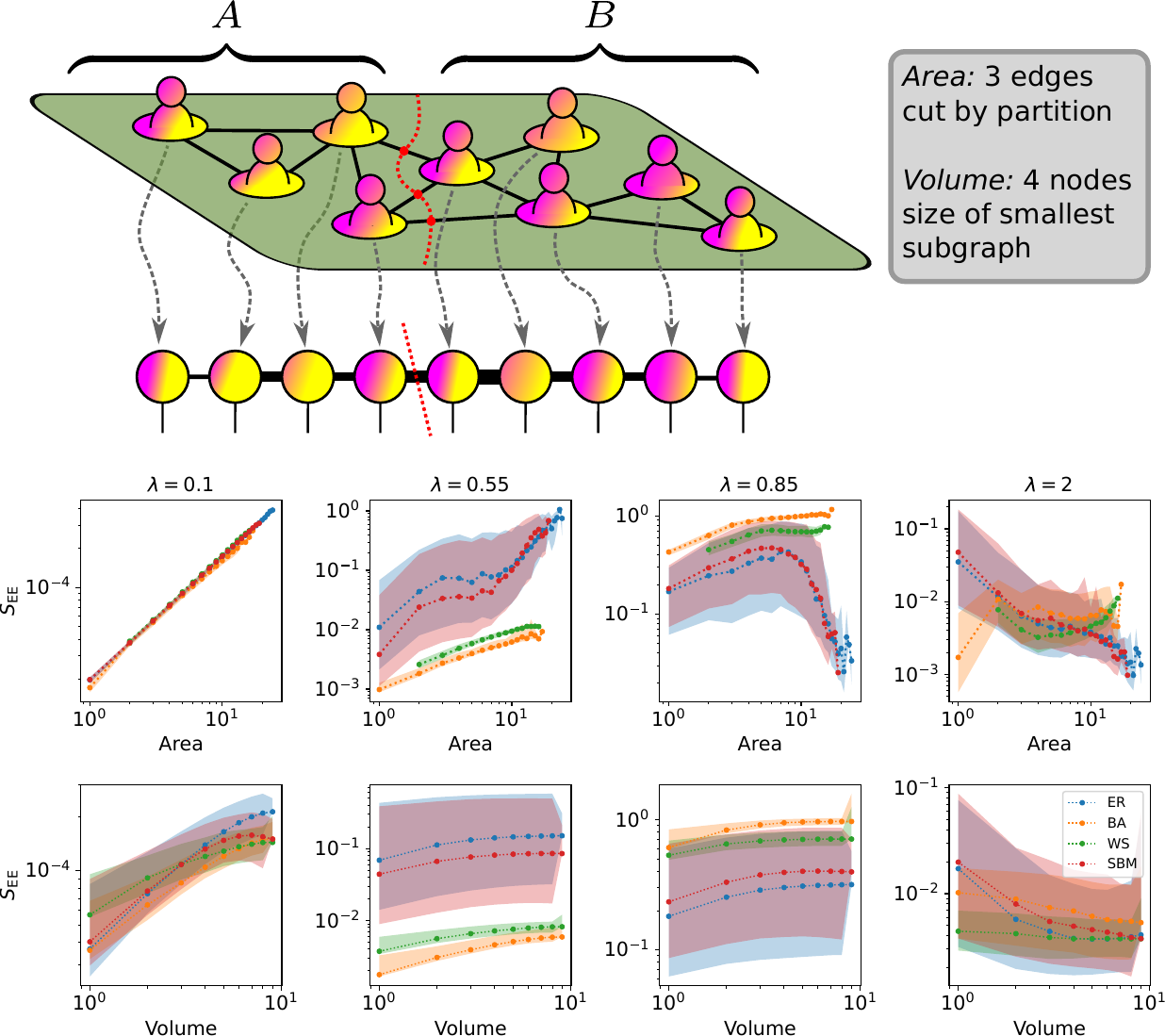}
    \caption{Each bond of the MPS represents a partition of the graph into two subgraphs (A and B). Here we investigate the entanglement entropy of such partitions, versus the number of edges which connect the two subgraphs (area, top row) and the size of the smallest subgraph (volume, bottom row). Each column corresponds to a different value of $\lambda$ and all results are averaged over 500 graphs generated by the indicated algorithm. The shaded areas mark the upper and lower standard deviations per graph type.}
    \label{fig:area_law_plot}
\end{figure}

As illustrated in Figure~\ref{fig:area_law_plot}, each bond of the MPS represents a partition of the graph into two subgraphs. The correlations between the two subgraphs are represented by the singular values within the bond. Here, we wish to investigate how the entanglement entropy depends on the number of edges between the two subgraphs of the partition (which we interpret as the \textit{area} of the partition), and how it depends on the size of the smallest subgraph (which we interpret as the \textit{volume} of the partition). 

The plots in Figure~\ref{fig:area_law_plot} give an indication of the functional dependence of $S_{\rm EE}$ with area (top row) and volume (bottom row). We see that for small $\lambda$, all studied graph generators show a clear correlation with the area (top left graph). Fitting a power law reveals an exponent close to 0.94 for all graph types. While, generally the entanglement entropy also increases with the volume of the partition, the relation is not so clear as with the area, which implies an area law holds for the network $\epsilon$-SIS model in the inactive phase. When $\lambda = 0.55$, around the critical value indicated by maximal variance, the entanglement entropy increases more rapidly with area for the ER and SBM networks, while BA and WS networks show power law increase in $S_{\rm EE}$ with exponents fitted to around $0.7$. For larger values of $\lambda$, the entanglement entropy starts to decrease again when the partition cuts a large number of edges. This indicates that regions with little connections to the remainder of the graph are more highly correlated. A possible explanation could be that for partitions with large areas, there are also a large number of different ways for the infection to spread across the partition, such that the correlation across the edges cut by the partition is less important. The dependence of $S_{\rm EE}$ on the volume of the partition seems to be only relevant for very small volume partitions, showing no significant dependence for larger volumes in the middle two columns. In the endemic phase, the small volume of the partition seems to imply a larger correlation with the rest of the system.  

\subsubsection{Influence of the bond dimension and the ordering on the computation time}

 \begin{figure}
     \centering
     \includegraphics[width=\linewidth]{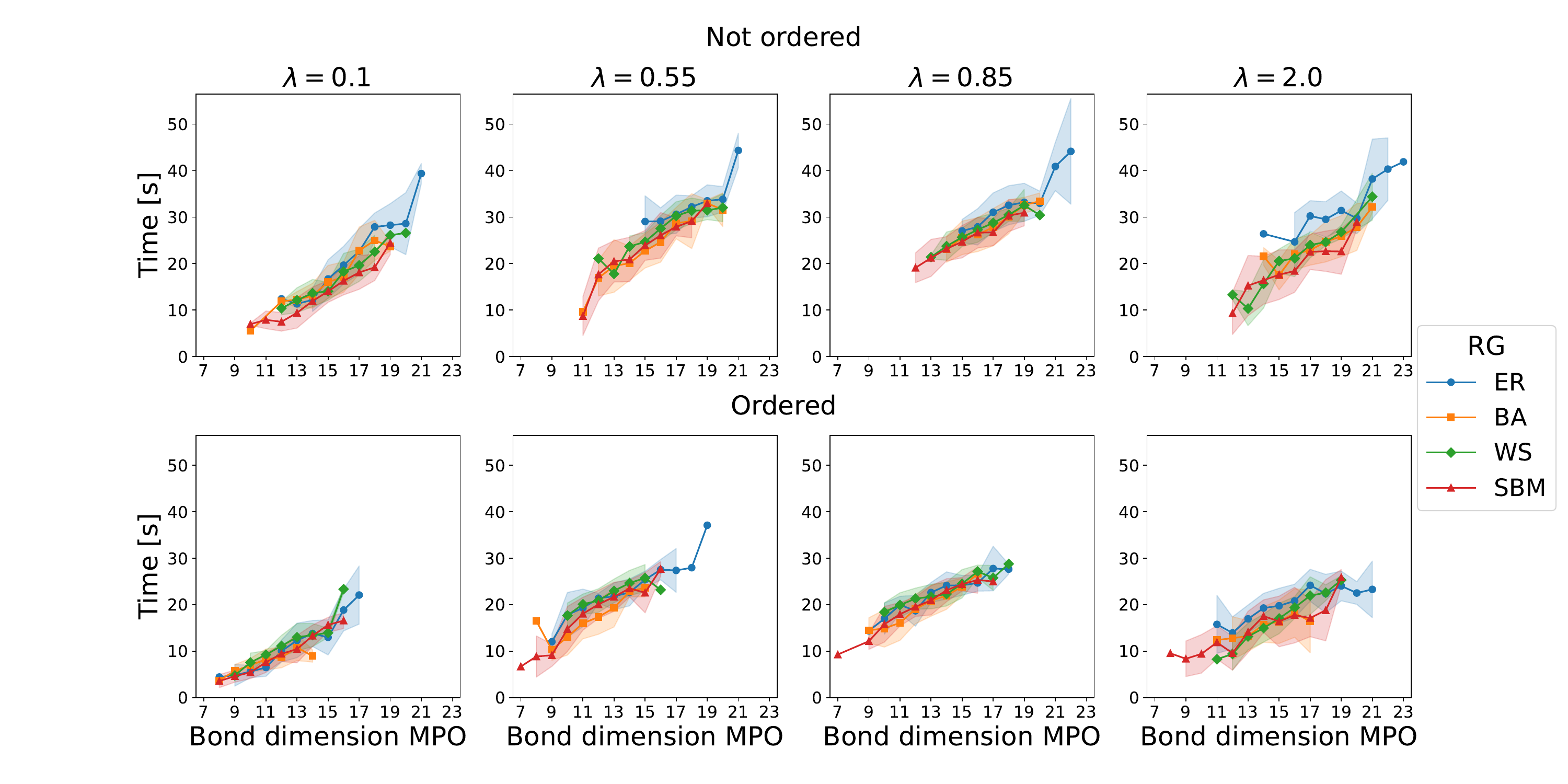}
     \caption{Computation time of the DMRG algorithm for $\lambda \in [0.1,0.55,0.85,2.0]$ as a function of the maximal bond dimension of the MPO for various random graph generators. The top row are the results for unordered MPOs, while the bottom rows has the same graphs first ordered according to the first Fiedler vector.}
     \label{fig:bonddimmpo}
 \end{figure}
 
The bond dimension of the MPO will influence the efficiency of the DMRG algorithm and this in turn is influenced by the ordering of nodes in the MPO/MPS Ansatz. In Figure~\ref{fig:bonddimmpo} we study the relationship between the computational efficiency and the bond dimension of the MPO for both unordered and ordered random networks. Here the computation time of the DMRG algorithm is tracked for 100 random graphs per type (ER, BA, WS and SBM) with the same graph parameters as in Figure~\ref{fig:meanfield_comparison} and for $\lambda \in [0.1,0.55,0.85,2.0]$ (we also take $\chi_{\rm max} = 128$ and $\delta = 10^{-10}$). For each random graph the bond dimension of the MPO is computed and grouped, after which the average and standard deviation are calculated. As for some MPO bond dimensions there is only one graph, these points do not have an error margin. In the top row, the graphs are not ordered according to the Fiedler vector, whereas in the bottom row the same graphs are ordered as described above. Note that both the computation time and the bond dimension of the MPOs decrease after reordering. The computation time of the DMRG scales (almost linear) with the bond dimension of the MPO. The value of $\lambda$ influences both the bond dimension and the computation time, but there does not seem to be a difference between different random graph generators within the error margins. This indicates that not graph topology, but rather the MPO bond dimension is an indicator of the algorithm's efficiency.

\subsubsection{Accuracy and efficiency comparison with exact methods}

\begin{figure*}
    \centering
    \includegraphics[width=0.8\textwidth]{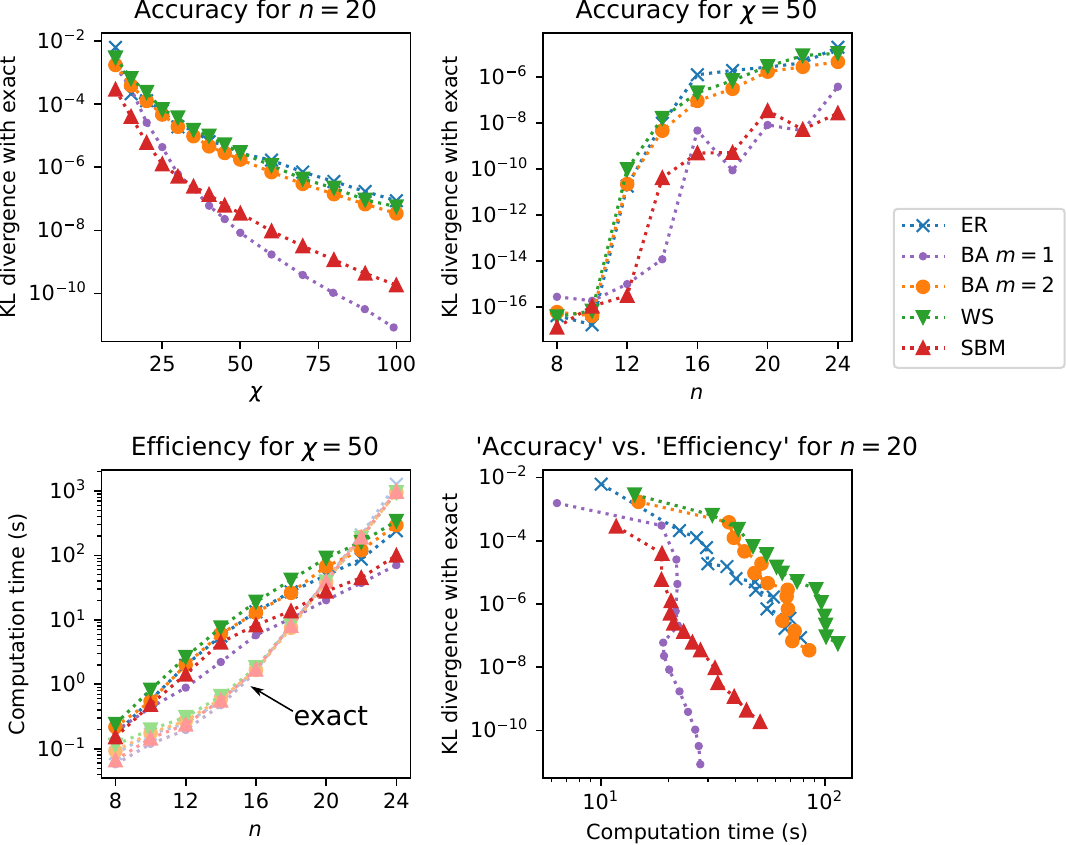}
    \caption{Accuracy and efficiency of the MPS methods compared to the exact distribution for five different random graph generating algorithms (ER for Erd\"os-Renyi, BA for Barab\'asi-Albert with $m=1,2$ new connections for each node, WS for Watts-Strogatz and SBM for the stochastic block model with 2 communities).}
    \label{fig:accuracyvsefficiency}
\end{figure*}

An advantage of the MPS representation over mean-field methods is that it gives an approximation of the full $2^n$-dimensional probability vector, retaining the relevant correlations in the system. The truncation of singular values provides a flexible trade-off between accuracy and efficiency, but how accurately can this truncation represent the probabilities over all $2^n$ network configurations? In Figure~\ref{fig:accuracyvsefficiency}, we compare the computational cost and accuracy of the MPS with that of exact diagonalization methods. The accuracy is now measured by computing the Kullback-Leibler (KL) divergence between the exact $2^n$ distribution and the same distribution computed from the MPS approximation. 

Here we investigate five different types of random graph generators: Erd\"os-Renyi (ER) graphs with $p=3/n$, Barab\'asi-Albert (BA) graphs both with $m=1$  (BA trees) and $m=2$, Watts-Strogatz (WS) graphs with $k=4, p = 0.25$ and the stochastic block model (SBM) with two equally sized clusters with internal link probability $0.5$ and external link probability $0.02$. Each data point in the figure represents the average over 100 graphs. We see that the BA trees (with $m=1$) and the SBM graphs are more accurately represented than the other networks, however, all network types show small KL divergence ($<10^{-5}$) with moderately small bond dimensions ($\chi =  50$). 

The efficiency is expressed in terms of computation time, where all computations were performed on the same machine to allow for fair comparison. We see in the bottom left panel of Figure~\ref{fig:accuracyvsefficiency} that for small network sizes, the highly optimized exact methods are faster, but around $n \sim 18-22$ the MPS overtakes the exact methods in terms of speed. The important factor here is scaling. While the exact methods scale exponentially as $d^n$, the MPS exhibits a substantially slower rate of increase in computational cost. This allows us to find accurate approximations of the full steady-state distribution for systems which are significantly larger than feasible with exact methods.

The bottom right panel of Figure~\ref{fig:accuracyvsefficiency} shows the accuracy in terms of the KL divergences between the MPS and the exact diagonalization, and the efficiency in terms of the computation time. Each data point in this plane represents the accuracy and cost of obtaining the MPS at a given fixed $\chi_{\rm max}$, where increasing $\chi_{\rm max}$ generally moves the line from top left to bottom right. As expected, most graphs see an increase in both computation time and accuracy with increasing bond maximal dimension. It is interesting to note that for BA trees, increasing the bond dimension does not necessarily increase the computation time. This is because in this case, even though a single DMRG sweep is faster with a smaller bond dimension, the algorithm will need more sweeps to converge when the bond dimension is too low to accurately represent the true state.

\begin{figure}
    \centering
    \includegraphics[width = 0.7\columnwidth]{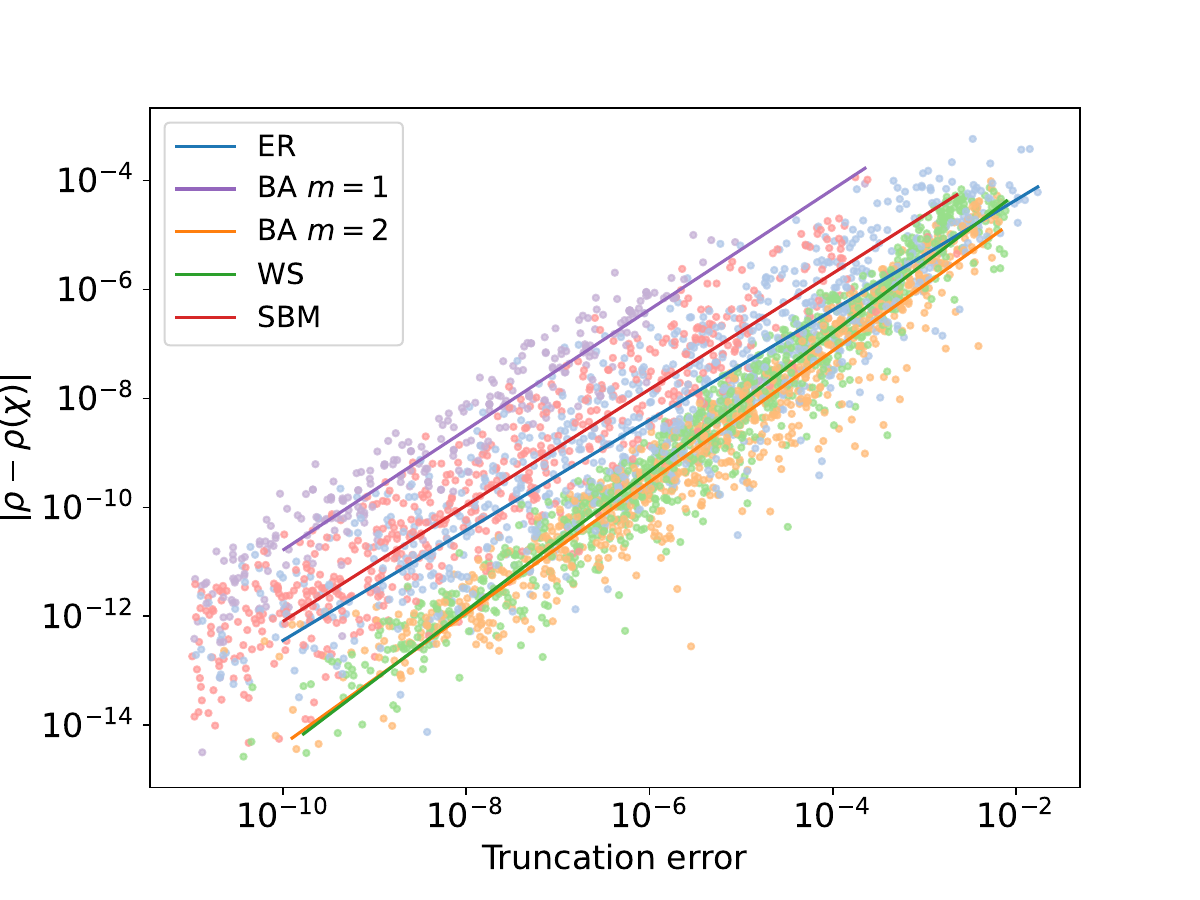}
    \caption{The error in density of the MPS representation versus the singular value truncation error, i.e. the size of discarded singular values. Different colors correspond to different graph generating algorithms. The lines are a power-law fits to the data.}
    \label{fig:truncation_error}
\end{figure}

Finally, Figure~\ref{fig:truncation_error} explores how the truncation of the singular value spectrum affects the accuracy of the representation. Each dot represents the difference between the density of infected nodes $\rho$ computed with an MPS of maximal bond dimension $\chi$ and the density of the exact solution. The truncation error represents the total magnitude of the discarded singular values when the MPS has converged. The different graph generators of Figure~\ref{fig:accuracyvsefficiency} are grouped together by color. The solid lines are power law fits of the accuracy $|\rho-\rho(\chi)|$ as a function of the truncation error. The exponents were found to be between $1.01$ (for ER networks) and $1.27$ (for the WS networks). Hence, the amount of discarded singular values can be used as a measure of the accuracy of the approximation, where a decrease in truncation error leads to an almost linear increase in accuracy.

\subsection{MPS representations of a real-world inspired network}

\begin{figure}[!t]
    \centering
    \includegraphics[width = \textwidth]{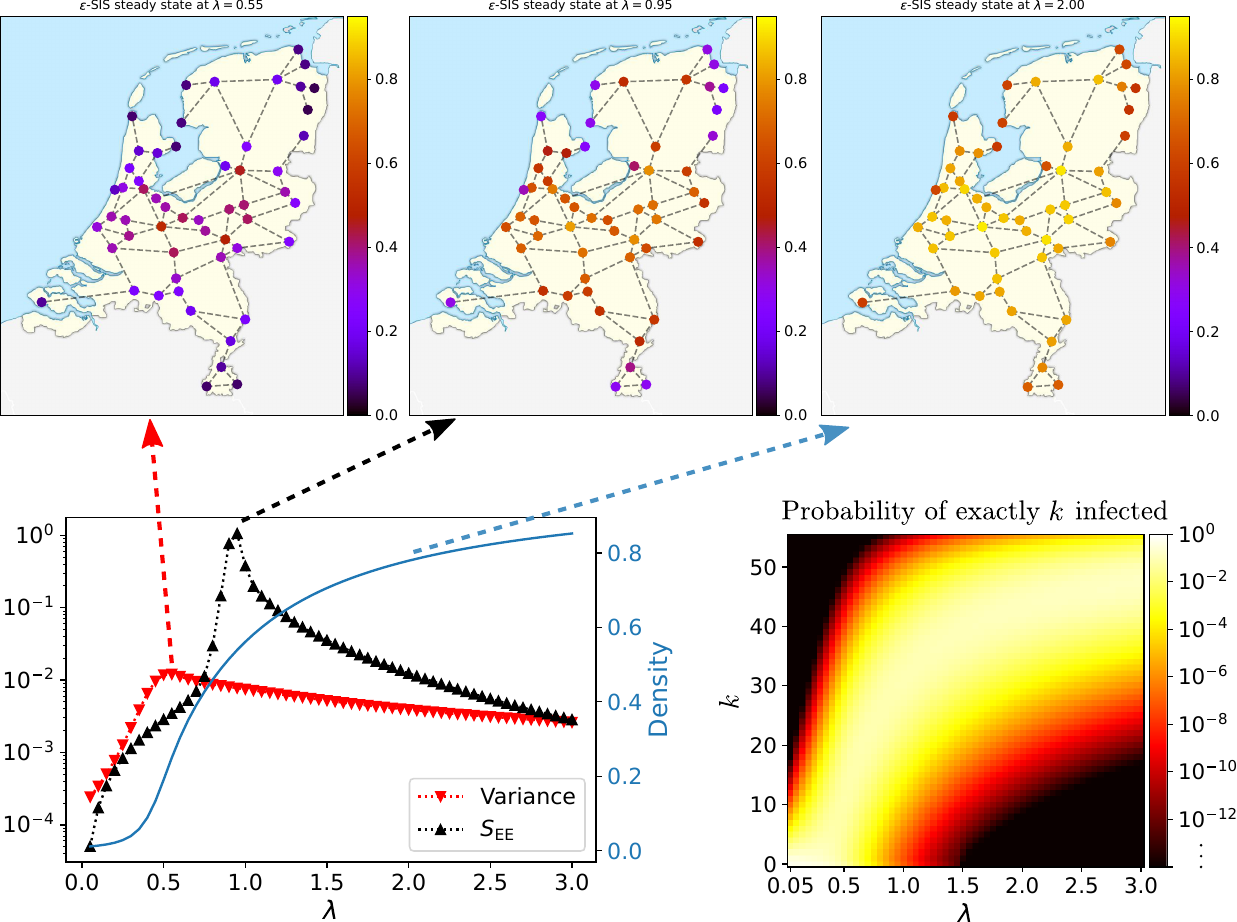} 
    \caption{The $\epsilon$-SIS process on a simplified version of the Dutch railway network with 55 nodes and 83 edges. Here, the spontaneous infection rate $\epsilon = 10^{-2}$ and recovery rate is normalized to unity $\gamma = 1$. Top maps show the individual nodes infected expectation values at three different values for $\lambda$. Bottom left plot shows the density of infected nodes in blue, the variance in density in red and the entanglement entropy in black. Bottom right shows the probability of having exactly $k$ infected nodes in the network, as computed from the MPS representation.}
    \label{fig:NL_trains}
\end{figure}

To illustrate the accuracy of the MPS method with a real-world example intractable with exact methods, we consider a simplified version of the Dutch railway network with 55 nodes and 83 edges. We compare the MPS results with Monte Carlo simulation and (second order) mean-field theory. 
The results for the node-base expectation values are shown for three values of $\lambda$ in the top panels of Figure~\ref{fig:NL_trains}. The bottom left panel shows the density, the variance in density and the entanglement entropy $S_{\rm EE}$ across the middle bond of the MPS. As before, the entanglement entropy peaks on the endemic side of the transition, whereas the variance peaks on the inactive side of the transition. 

\begin{figure}
    \includegraphics[width = \textwidth]{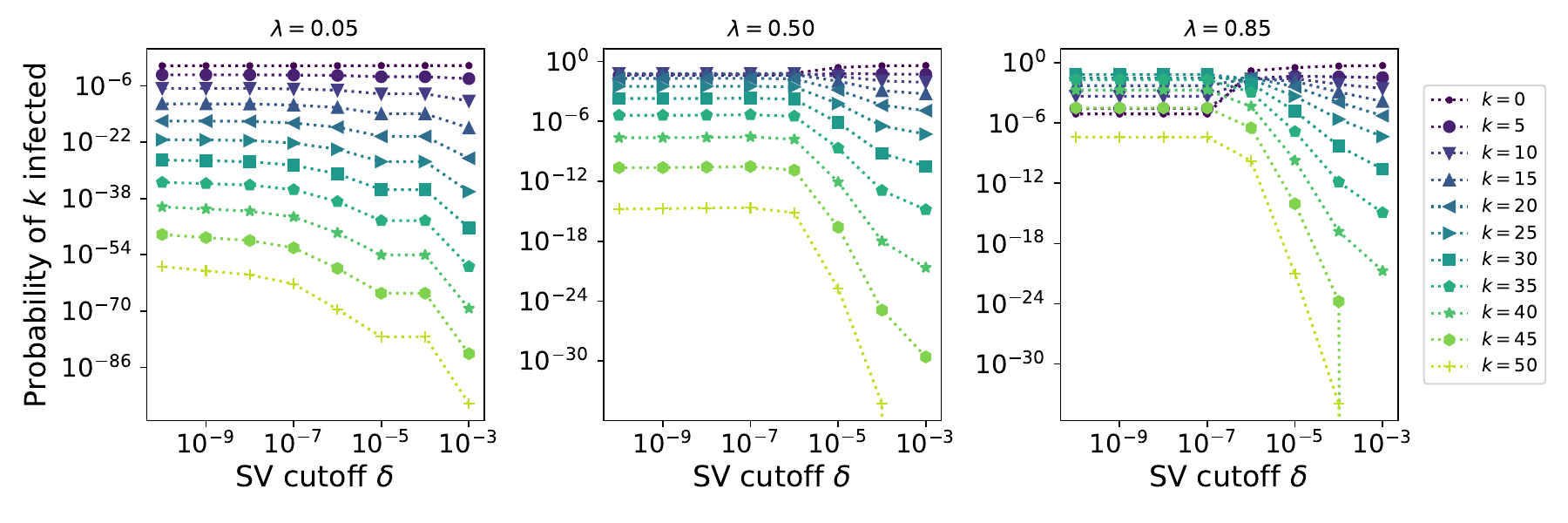}
    \caption{The dependence on singular value cutoff $\delta$ of the probability of observing exactly $k$ infected in the train network for various $k$ and three values of $\lambda$. }
    \label{fig:cutoff_dependence}
\end{figure}

We furthermore show the probability of having exactly $k$ infected nodes in the system, as a function of both $k$ and the transmission rate $\lambda$ in the bottom right panel of Figure~\ref{fig:NL_trains}. We compute this by first constructing an MPO which projects the MPS on all configurations with exactly $k$ infected, followed by computing the expectation value of the MPO. This MPO has bond dimension $\chi_{\rm mpo} = {\rm min} (k+1, N-k +1)$, as when $k>N/2$ an MPO with bond dimension $N-k+1$ can be made which projects on all configurations with $N-k$ susceptible nodes. For contrast reasons, whenever the probabilities become smaller that $10^{-14}$, we display it as black, whereas in reality, some combinations of $k$ and $\lambda$ give probabilities of occurring as low as $10^{-56}$. When the expected probabilities are this low, one may wonder how accurate these results are. This can be tested by running the MPS with various truncation cutoffs $\delta$ and studying the stability of the predictions. Figure~\ref{fig:cutoff_dependence} shows that while the extremely low probability events as still quite sensitive to the cutoff threshold, probabilities of the order of $10^{-14}-10^{-16}$ are quite stable when lowering the singular value threshold below $\delta = 10^{-7}$. 
This shows that the MPS is not only good for computing averaged quantities and variances, but can also be used to efficiently obtain information of events with very low probability of occurring and hence constitutes a useful tool on obtaining accurate statistics of rare events.

\begin{figure}[!t]
    \includegraphics[width = 0.75\textwidth]{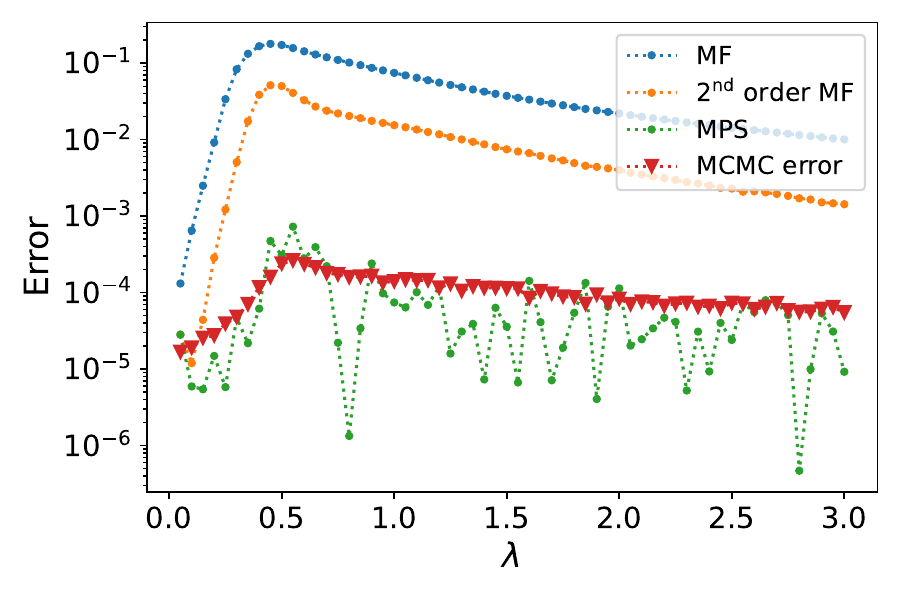}
    \caption{The difference in densities comparing Markov Chain Monte Carlo (MCMC) results with mean-field method (MF), second order mean-field ($2^{\rm nd}$ order MF) and the MPS method. The MCMC is averaged over $2 \cdot 10^6$ data points and the MCMC error estimated by the batch means method are indicated in red, implying that the MPS error is comparable to the MCMC accuracy and hence the MPS error is an upper bound on its accuracy.}
    \label{fig:comparison_with_MF_MC}
\end{figure}

The MPS results on the train network were obtained by setting a maximal bond dimension of $\chi_{\rm max} = 250$ and a singular value cut-off threshold of $\delta = 10^{-12}$. As illustrated in Figure~\ref{fig:comparison_with_MF_MC}, we managed to get very accurate results. There, we compare the difference in density of infected with the result obtained from $2\cdot 10^6$ MCMC runs, obtained by a standard implementation of Gillespie's algorithm on a network. The mean-field results show quite a deviation to the MCMC averages, whereas the MPS has an accuracy comparable to the MCMC errors shown in red. This implies the MPS result is accurate within the error bar of the MCMC, averaged over 2 million runs. Hence the error shown in green in Figure~\ref{fig:comparison_with_MF_MC} is a conservative estimate of the MPS accuracy.

\section{Conclusion and discussion}
\label{sec:conclusions}

In summary, we have demonstrated here that the MPS method may be used to provide accurate and efficient descriptions of the high-dimensional probability distributions describing the steady-state vector of the $\epsilon$-SIS model on a network. The MPS representation gives improved accuracy compared to mean-field methods for moderate bond dimensions and provides a systematic way to tune the accuracy of the approximation by increasing the bond dimensions. Moreover, the MPS gives a low-dimensional approximation for the probability of the systems being in any of its $2^n$ configurations, which can be used to extract meaningful information on the probabilities of rare events in an efficient way. 

To provide a quantitative measure for the compressibility of the state vector of the system, we have defined the notion of entanglement entropy analogously to the quantum information measure used in many-body quantum systems. This gives an indication of the compressibility of the system, and it follows the intuition that more complex states are harder to compress accurately. We have seen that for randomly generated networks of various types, the entanglement entropy converges as the bond dimension is increased, indicating that at this point the dimensionality of the representation is large enough to capture all relevant (linear combinations of) basis states. It is exactly for those bond dimensions where the entanglement entropy is converged that an increase in the accuracy of the MPS approximation is observed. 

Furthermore, we have studied the entanglement entropy as a function of the transmission rate $\lambda$ and observed that it peaks close to criticality, but slightly on the endemic side of the phase transition. This is in line with the expectation that the maximally complex states are those at the `edge of chaos', where here the chaotic phase corresponds to the disordered phase. It would be very interesting to explore the connection between the entanglement entropy and `the edge of chaos' further in other types of systems and see whether this can be used as a good measure of the compressibility in other systems as well.      

We have also studied in detail an MPS representation of a railway network with 55 nodes and 83 edges, which makes exact computation infeasible. The MPS can provide accurate statistics of the exact Markovian model, improving on the second order mean-field theory by several orders of magnitude.\footnote{Note, however, that more sophisticated moment closure techniques exist than the one utilized in this paper \cite{kuehn2024preserving}. It would be interesting to compare the MPS methods against the optimal choice for moment closures at the second order.} Furthermore, it may be used to provide insight into the statistics of rare events without additional computational cost, improving upon Monte Carlo sampling methods whose computational cost scales inversely proportional to the likelihood of the rare event~\cite{merbis2023efficient}.

It would be interesting to further explore generalizations of the methods presented here which could go in various directions. One could adapt and modify the process by including models with more compartments (such as susceptible-infected-recovered (SIR), or SIRS, susceptible-exposed-infected-recovered (SEIR))~\cite{kermack1932contributions,kendall1956deterministic}, or other stochastic models of complex systems, such as opinion dynamics, polarization or segregation models~\cite{castellano2009statistical,dall2008statistical}. An interesting way to be able to increase the usefulness of this method, which would allow for larger networks, is to combine the tensor network approach with network renormalization techniques~\cite{villegas2023laplacian,garuccio2023multiscale}, or to first build meta-population models~\cite{keymer2000extinction,colizza2008epidemic} where each node represents a groups of individuals.
Other promising avenues leading to `mixture models' would be to combine TN techniques with message-passing (or believe propagation) algorithms \cite{cantwell2019message} or with higher-order interactions, leading to non-linear infection kernels \cite{st2021universal}. Recently, a loop series expansion for believe propagation algorithms with tensor networks was proposed in \cite{evenbly2024loop} to systematically improve in accuracy. It would be very interesting to apply these methods to stochastic models on complex networks.

Another direction for future work is to study the time-evolution from a given initial state using time-evolving block decimation (TEBD) algorithms~\cite{vidal2003efficient}. In this framework, the time evolution can be mapped to performing a non-unitary quantum computation. Finally, an interesting avenue is to adapt these methods to other tensor network architectures, such as the hierarchical Tree tensor networks~\cite{silvi2019tensor}, which could be a more efficient in the presence of long-range correlation, and Projected Entangled-Pair States (PEPS)~\cite{verstraete2004, nishio2004, orus2014practical, verstraete2008matrix, vlaar21} for higher-dimensional lattice systems.

\ack
We wish to thank Piet Van Mieghem, Mauricio del Razo, Miguel A. Mu\~noz, Manlio De Domenico and Simone Montangero for stimulating discussions. All results of this paper are produced using the \href{https://github.com/wmerbis/stochasticTN/}{\texttt{stochasticTN}} package developed by the authors and made publicly available through GitHub. PC acknowledges support from the European Research Council (ERC) under the European Union’s Horizon 2020 research and innovation programme
(Grant Agreement No. 101001604).

\section*{References}
\providecommand{\newblock}{}

\end{document}